\documentclass[pre,eqsecnum,twocolumn, showpacs,nofootinbib]{revtex4}
\global\arraycolsep=2pt
\usepackage{stmaryrd}
\usepackage{amsmath}
\usepackage{amssymb}
\usepackage{graphicx}
\usepackage{textcomp}
\usepackage{calrsfs}
\usepackage{bm}

\newcommand{\be}{\begin{equation}}                                                
\newcommand{\ee}{\end{equation}}
\newcommand{\ben}{\begin{equation*}}                                              
\newcommand{\een}{\end{equation*}}
\newcommand{\bea}{\begin{eqnarray}}   
\newcommand{\eea}{\end{eqnarray}}
\newcommand{\nn}{\nonumber}
\begin{document} \scrollmode

\title{Investigations of the torque anomaly in an annular sector.\\
  I. Global calculations, scalar case}

\date{\today}

\author{Kimball A. Milton}\email{milton@nhn.ou.edu}
\author{Fardin Kheirandish}\email{fkheirandish@yahoo.com}
\altaffiliation{Permanent address: 
Department of Physics, Faculty of Science, University of Isfahan,
Hezar-Jarib St., 81746-73441,
Isfahan, Iran}
\author{Prachi Parashar}\email{prachi@nhn.ou.edu}
\author{E. K.  Abalo}\email{abalo@nhn.ou.edu}
\affiliation{Homer L. Dodge  Department of
Physics and Astronomy, The University of Oklahoma, Norman, OK 73019-2053}
\author{Stephen A. Fulling}\email{fulling@math.tamu.edu}
 \author{Jeffrey D. Bouas}\email{jdbouas@gmail.com}
 \author{Hamilton Carter}\email{hcarter333@neo.tamu.edu}
\affiliation{Departments of Mathematics and Physics, Texas A\&M University,
College Station, TX 77843-3368}
 \author{Klaus Kirsten} \email{Klaus_Kirsten@baylor.edu}
 \affiliation{Department of Mathematics, Baylor University, One 
Bear Place, Waco, TX 76798-7328}
 
 \begin{abstract}
In an attempt to understand a recently discovered torque anomaly 
in quantum field theory with  boundaries, we calculate the 
Casimir energy and torque of a scalar field subject to Dirichlet 
boundary conditions on an annular sector  defined by two coaxial 
cylinders intercut by two planes through the axis.  
 In this model the particularly troublesome divergence at the 
cylinder axis  does not appear, but new divergences associated 
with the curved boundaries are introduced.
All the
divergences associated with the volume, the surface area, the 
corners, and the curvature  are regulated by 
point separation either in the direction of the axis of the 
cylinder or in the (Euclidean) time;  the full divergence 
structure is isolated, and the remaining finite energy and torque 
are extracted.  Formally, only the regulator based on axis 
splitting yields the expected balance between energy and torque.  
 Because of the logarithmic curvature divergences, 
there is an ambiguity in the linear dependence of the energy on the 
wedge angle; if the terms constant and linear  in this angle are 
removed by a process of renormalization, the expected 
torque-energy balance is preserved. 

  \end{abstract}

\maketitle

\section{Introduction}
Recently, evidence was presented \cite{fulling12} that the expected
relation between energy and torque may not be satisfied by quantum
vacuum energy. 
  This conclusion is hard to accept, since the energy--torque balance
formally  follows from the general
underlying variational principle \cite{dowker13}. 
  Specifically, 
Ref.~\cite{fulling12} considers the vacuum expectation values
of the energy-momentum tensor for a wedge, for both a conformally coupled 
 scalar field,
where the wedge surfaces are Dirichlet boundaries, and for electromagnetism,
where the boundaries are perfect conductors.  These expectation values
were evaluated many years ago by Dowker and Kennedy for the scalar
case \cite{Dowker:1978md} and then more generally by
 Deutsch and Candelas  \cite{deutsch}.
(See also Refs.~\cite{brevik}  and \cite{saharian}.) 
  Those papers presented
calculations of the local stress tensor in the wedge geometry;
  if the $\langle
T^{00}\rangle$ component is integrated over the region inside the
wedge, or if $\langle T_{\theta\theta}\rangle$ 
is integrated over one of the bounding
planes of the wedge, divergences are encountered because of the singularity
of the field at the axis of the wedge.  
In Ref.~\cite{fulling12} integration
is therefore extended over a finite range of radial distances from the
apex, and then it is found that the torque is not equal to the negative
derivative of the energy with respect to the opening angle of the wedge.
 Because the stress tensors of the confomally invariant fields are 
completely finite in the radial range concerned,
 this result might be considered 
quite different from the pressure anomaly found 
earlier \cite{Estrada:2012yn} for orthogonal plane boundaries, which can be 
blamed on a physically faulty regularization prescription.  However, to
extract the finite stress tensor, regularization of divergent quantities
is required, so an issue arises that must be explored further.

\begin{figure}[tb]
  \begin{center}
    \includegraphics[width=2in]{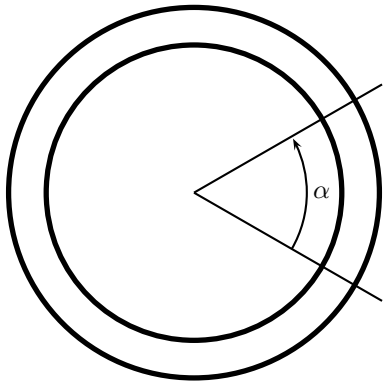}   
    \caption{\label{fig-ann} Two radial Dirichlet  plates, forming 
a dihedral angle $\alpha$, intersecting
two coaxial Dirichlet cylinders of radius $a$ and $b$, forming an
 annular region.  If the plane ribbons between the cylinders  are imagined
as free to slide in the circumferential, $\theta$, direction, 
maintaining contact with the
cylindrical walls, this constitutes an annular piston.  In this paper
we are considering the interior sector alone; the exterior results are
obtained by the replacement $\alpha\to 2\pi-\alpha$.}
  \end{center}
\end{figure}

 It was immediately objected that  this calculation does not describe a complete and 
physically acceptable model system.
 If the inner radius is taken to zero, one encounters the 
divergences at the axis; by subtracting infinities one 
 can prove any result imaginable.
 If the reflecting wedge walls are simply truncated at a finite 
radius, one has 
 a pair of finite, nonintersecting planes, requiring
 a different, and much harder, calculation.
Instead, one can insert  reflecting cylindrical  boundaries at the 
inner and outer radii, producing a truncated sector as in 
 Fig.~\ref{fig-ann}.
 (It transpires that the outer cylinder is not very 
important, because the expectation values
of the stress tensor fall off rapidly with 
radius.  One could take the outer radius to infinity, but in this 
paper we prefer to keep it finite so that all terms in the 
 regularized total energy will be finite.)
 This model is the most promising to study.  Unfortunately, it 
introduces new divergences associated with the curvature of the 
bounding cylinders.  Worse, some of these divergences are 
logarithmic 
(or would be associated with a pole of the zeta function), thereby 
creating an inherent ambiguity in the finite parts.  As we shall see,
this ambiguity is intimately connected to the existence of a torque
anomaly.
 
Dowker \cite{dowker13} has outlined how the seemingly paradoxical 
 torque anomaly might be resolved by solving the annular 
sector model introduced in the previous paragraph.
 The energy density and torque density associated with boundary 
regions near the 
cylinders, at finite radii, may display a precisely compensating 
imbalance, so that the annular system is nonanomalous.
If this imbalance 
does not approach zero as the inner radius shrinks, its effect will 
be incorrectly lost in the calculation for the full wedge, in which 
there is no separate contribution from the apex of the wedge per~se.

In  this paper
  we examine a Dirichlet  wedge intersected with a pair
of coaxial, circular Dirichlet cylinders, as shown in Fig.~\ref{fig-ann}.  
Thus an ``annular piston'' is realized,
in which boundary motion takes place only inside
  the finite region defined between the concentric cylinders.
  That region consists of two ``annular sectors'' separated by the 
  radial planes. 
 In this paper we consider the total energy and torque for the 
scalar field in an annular sector, leaving local and 
electromagnetic calculations for future papers.
   We find the expected relation
between the formal, unregulated expression for the torque on the radial planes
 and that for the interior  energy, both of which are formally divergent.
These expressions may be regulated by a regulator which does not depend on
the opening angle of the wedge, in which case
we might expect that  the divergent (as the regulator
goes to zero) and finite parts obey the expected energy-momentum balance.
However, this turns out to be not quite the case:
  First, if all terms are taken
seriously, only a neutral regulator (point separation in a direction not
involved in the components of the stress tensor involved,
  in this case $z$ or $r$, 
 the latter being less desirable because there is a boundary in that
direction) gives the expected balance between energy and torque;
 this is precisely as expected from Ref.~\cite{Estrada:2012yn}.
 However, because
of the logarithmic curvature divergences, there is an ambiguity in
the linear dependence on the wedge angle in both
the divergent and finite parts of the energy; if this arbitrary linear
dependence is removed by a process of ``renormalization,'' all divergences and
anomalies disappear, and the physical torque is the negative derivative
with respect  to the wedge angle of the physical energy, in agreement with
previous considerations of the annular piston~\cite{Milton:2009bz}.

\section{Torque and energy on wedge intersected by 
coxial cylinders}
\label{sec:ap}

We  compute the torque and energy for the annular sector defined
in Fig.~\ref{fig-ann}, starting from the canonical flat-space stress tensor
for a massless scalar field,
\be
T^{\mu\nu}=\partial^{\mu}\phi\partial^\nu\phi-\frac12g^{\mu\nu}
\partial_\lambda
\phi\partial^\lambda\phi,
\ee
in terms of the metric
\be
g_{\mu\nu}=\mbox{diag}(-1,1,\rho^2,1).
\ee
Thus, the angular-angular part of the stress tensor is
\be
T^{\theta}{}_{\theta}=\frac12\left(\partial^0\phi\partial^0\phi-\partial_\rho
\phi\partial_\rho\phi+\frac1{\rho^2}\partial_\theta\phi\partial_\theta\phi-
\partial_z\phi\partial_z\phi\right).
\ee
Here we adopt cylindrical coordinates, with the $z$ axis along the cylinder
axis. Now to get the quantum vacuum stress, we replace
\be
\phi(x)\phi(x')\to\langle \phi(x)\phi(x')\rangle=\frac1i G(x,x'),
\ee
in terms of the Feynman or causal Green's function.  For the geometry
considered we can represent the Green's function as
\bea
G(x,x')&=&\int\frac{d\omega}{2\pi}e^{-i\omega(t-t')}\int\frac{dk_z}{2\pi}
e^{i k_z(z-z')}\nn\\
&&\quad\times\sum_\nu\Theta_\nu(\theta)\Theta_\nu^*(\theta')
g_\nu(\rho,\rho').\label{wedgegf}
\eea
Here the eigenfunctions in $\theta$ and the corresponding eigenvalues
$\nu$ are given by 
\be
\left[-\frac{\partial^2}{\partial\theta^2}+v(\theta)\right]\Theta_\nu(\theta)
=\nu^2\Theta_\nu(\theta),
\ee
where the potential $v(\theta)$ represents the plane ribbons, so
$v(\theta)=0$ within the sector.
From this, we can integrate over the wedge surface at $\theta=\alpha$ to
compute the torque on the plane at $\theta=\alpha$ 
per unit length (in the $z$ direction),
\bea
\tau&=&\int_0^\infty d\rho\,\rho\langle T^{\theta}{}_{\theta}\rangle
(\rho,\alpha)
 \nn\\
&=&\int_0^\infty d\rho\,\rho \frac1{2i}\int\frac{d\omega}{2\pi}
\int\frac{dk_z}{2\pi}\sum_\nu
\frac1{\rho^2}\\
&&\times( \partial_\theta\partial_{\theta'}+\nu^2)
\Theta_\nu(\theta)\Theta_\nu^*(\theta') g_\nu(\rho,\rho')
\bigg|_{\theta'=\theta=\alpha, \rho'=\rho}.\nn
\eea

Let us simplify the following discussion by considering Dirichlet wedge
surfaces, where the eigenvalues are explicit:
\be
\nu=mp,\quad  p=\frac\pi \alpha,
\ee
and the eigenfunctions are explicitly
\be
\Theta_\nu(\theta)=\sqrt{\frac2\alpha}\sin mp\theta.
\ee
Then the torque due to the field fluctuations interior to the annular sector is
\be
\tau=\frac1{2\pi\alpha}\sum_{m=1}^\infty (mp)^2\int_0^\infty d\kappa\,\kappa
\int_a^b \frac{d\rho}\rho g_{mp}(\rho,\rho),\label{torqueform}
\ee
where we have made a Euclidean rotation, $\omega\to i\zeta$, and
adopted polar coordinates with $\kappa=\sqrt{\zeta^2+k^2_z}$.

The reduced Green's function in the interior of the annular sector,
$\rho\in[a,b]$, 
is
\bea
g_\nu(\rho,\rho')&=&I_\nu(\kappa \rho_<)K_\nu(\kappa \rho_>)\nn\\
&&\quad\mbox{}-I_\nu(\kappa \rho)I_\nu(\kappa \rho')
\frac{K_\nu(\kappa a)K_\nu(\kappa b)}{\Delta}\nn\\\
&&\quad\mbox{}-K_\nu(\kappa \rho)K_\nu(\kappa \rho')
\frac{I_\nu(\kappa a)I_\nu(\kappa b)}{\Delta}\nn\\\
&&\quad\mbox{}+[K_\nu(\kappa \rho)I_\nu(\kappa\rho')+I_\nu(\kappa\rho)
K_\nu(\kappa\rho')]\nn\\
&&\quad\quad\times
\frac{I_\nu(\kappa a)K_\nu(\kappa b)}{\Delta},
\eea
where the denominator is \be
\Delta=\Delta_\nu(\kappa a,\kappa b)=I_\nu(\kappa b)K_\nu(\kappa a)
-I_\nu(\kappa a)K_\nu(\kappa b).
\ee
Now in terms of $Z_\nu(z), \tilde Z_\nu(z)=I_\nu(z)$ or $e^{i\nu\pi}K_\nu(z)$,
the radial integrals may be evaluated by the following indefinite integral
\cite{prudnikov}:
\bea
&&\int\frac{dz}z Z_\nu(z)\tilde Z_\nu(z)=-\frac{Z_\nu(z)\tilde Z_\nu(z)}{2\nu}
\\
&&\!\!\!\!\!\!\!\mbox{}-\frac{z}{2\nu}\left(Z_{\nu-1}(z)
\frac\partial{\partial \nu}\tilde Z_\nu(z)-Z_{\nu}(z)\frac{\partial}{\partial
\nu}\tilde Z_{\nu-1}(z)\right).\nn
\eea
Consequently, the result of the radial integral in Eq.~(\ref{torqueform}) is found
by straightforward algebra:
\be
\int_a^b \frac{d\rho}\rho g_\nu(\rho,\rho)=\frac1{2\nu}\frac\partial{\partial
\nu}\ln\Delta_\nu(\kappa a,\kappa b).
\ee
 From this,  using 
\be
\frac{\partial}{\partial\nu}=-\frac\alpha\nu \frac{\partial}{\partial\alpha},
\label{dnisda}
\ee
we find
\be
\tau=-\frac{\partial}{\partial\alpha}\frac1{4\pi}\sum_{m=1}^\infty 
\int_0^\infty d\kappa\,
\kappa\ln\Delta_\nu(\kappa a,\kappa b).\label{torque2}
\ee
This exhibits the expected relation between torque and energy,
\be
\tau=-\frac{\partial}{\partial\alpha}\mathcal{E},\label{tauenergy}
\ee
 provided the energy per unit length is
\be
\mathcal{E}=\frac1{4\pi}\sum_{m=1}^\infty \int_0^\infty d\kappa\,
\kappa\ln\Delta_\nu(\kappa a,\kappa b).\label{enint}
\ee

In fact, it is easy to calculate the interior energy of the annular sector.
This may be computed from the integrated energy density, 
\be
\langle T^{00}\rangle=\frac1{2i}[\partial^0\partial^{\prime0}-(\partial^0)^2]
G(x,x')\big|_{x'=x},\label{veve}
\ee
which leads to the
general formula \cite{miltonbook}
\be 
E=\int (d\mathbf{r}) \int_{-\infty}^\infty \frac{d\omega}{2\pi i}
\,\omega^2 \mathcal{G}(\mathbf
{r,r};\omega),
\ee
in terms of the Fourier transform of the Green's function
\be
G(x,x')=\int_{-\infty}^\infty \frac{d\omega}{2\pi} 
e^{-i\omega(t-t')}\mathcal{G}(\mathbf{r,r};
\omega).
\ee
Here this reads for the energy per unit length
\be
\mathcal{E}=-\frac1{4\pi}\int_0^\infty d\kappa\,\kappa^3\int_a^b d\rho\,\rho
\sum_{m=1}^\infty g_\nu(\rho,\rho').\label{unregenergy}
\ee  
The radial integrals are straightforwardly evaluated
\cite{prudnikov}.  The result 
\be
\mathcal{E}=-\frac1{8\pi}\sum_{m=1}^\infty \int_0^\infty d\kappa\,
\kappa^2\frac\partial{\partial \kappa}\ln\Delta_\nu(\kappa a,\kappa b)
\ee
is indeed the energy shown in Eq.~(\ref{enint}), after integration by parts,
at this point purely formal.

Thus, we see at a formal level there is the correct balance between
energy and torque, Eq.~(\ref{tauenergy}).  
Of course, the expressions for the torque and the energy given here
are divergent. 
However, we might expect that as long as these integrals are regulated in 
a way that does not refer to the wedge angle $\alpha$, the divergent and finite
parts of the torque and the energy will satisfy the same balance.  
In Sec.~\ref{sec5}
we will investigate this by introducing a point-splitting cutoff.

\section{Conformal terms}
\label{sec:ct}
The above used the canonical stress tensor for the scalar field.  More
generally, the stress tensor is
\be
T^{\mu\nu}=\partial^\mu\phi\partial^\nu\phi-\frac12 g^{\mu\nu}\partial_\lambda
\phi\partial^\lambda\phi-\xi(\partial^\mu\partial^\nu
 -g^{\mu\nu}\partial^2)\phi^2,
\ee
where $\xi$ is an arbitrary parameter, and the corresponding term is identically
conserved.  The value of $\xi$ that makes conformal invariance manifest, 
 and that makes the ultraviolet behavior of the theory most
regular, is, in three
spatial dimensions, $\xi=\frac16$ \cite{Callan:1970ze, Milton:1972cc}.  
For example, when $\xi=1/6$,
the divergences in the energy density near a Dirichlet surface are reduced
 from $1/z^4$ to $1/z^3$, where $z$ is the distance from the 
surface \cite{Milton:2010qr}.  
However, this ambiguity is irrelevant when computing global quantities such as
the total energy or the total torque  \cite{Milton:2010qr}.  
The reason is the following:
Let us denote the conformal term in the stress tensor as
\be
\Delta T^{\mu\nu}=-\xi(\partial^\mu\partial^\nu-g^{\mu\nu}\partial^2)\phi^2.
\ee
For the torque contribution we need
\be
\Delta T^{\theta\theta}=-\xi\left((\partial^0)^2-\partial_z^2-\frac1\rho
\partial_\rho\rho\partial_\rho\right)\phi^2,
\ee
which for Dirichlet boundaries
vanishes on either plate, $\theta=0$ or $\theta=\alpha$.  This continues to be
true if we point split in a direction which respects the boundary conditions, 
that is,
in the $z$ direction or the $t$ direction.  For the energy contribution we have
\be
\Delta T^{00}=-\xi\nabla^2\phi^2,
\ee
which, when integrated over a region with Dirichlet boundaries, vanishes because
\be
\int_V(d\mathbf{r})\nabla^2\phi^2=2\oint_{\partial V}d\bm{\sigma}
\cdot\phi\bm{\nabla}\phi=0.
\ee
This again holds even with point splitting in a direction perpendicular 
to the normals of the  boundaries,
here in the $z$ or $t$ directions.  

\section{Point splitting}
\label{sec5}
We now redo the calculation in Sec.~\ref{sec:ap}.  In view of the discussion 
in Sec.~\ref{sec:ct},
we point split in the directions either of $t$ or $z$.  
After the Euclidean rotation, the torque per length is
\be
\tau=\frac12\int_a^b \frac{d\rho}\rho\int\frac{d\zeta\,dk}{(2\pi)^2}
e^{i\zeta t_E}e^{ikZ}
\frac2\alpha\sum_{m=1}^\infty (mp)^2g_{mp}(\rho,\rho),
\ee
in place of Eq.~(\ref{torqueform}), where $t_E=i(t-t')$ and 
$Z=z-z'$ 
are point-split regulator 
parameters, which are to be taken to zero.  Now let us write
\begin{subequations}
 \bea
\zeta&=&\kappa\cos\gamma,\quad k=\kappa\sin\gamma,\\
t_E&=&\delta\cos\phi,\quad Z=\delta\sin\phi,
\eea
\end{subequations}
where $\delta\to0$ and $\phi$ is an arbitrary angle.  When $\phi=0$
we are doing time splitting, while if $\phi=\pi/2$, we are splitting the
points at which the product of fields are evaluated at 
slightly different values of $z$.
Thus, the regulator exponent is
\be \zeta t_E+k Z=\kappa\delta\cos(\gamma-\phi).\ee
The regulated torque is, in fact, independent of $\phi$, and in place of 
Eqs.~(\ref{torqueform}) and (\ref{torque2}) we have
\bea
\tau&=&\frac1{2\pi\alpha}\int_a^b\frac{d\rho}\rho\int_0^\infty d\kappa\,\kappa
\, J_0(\kappa\delta)
\sum_{m=1}^\infty (mp)^2g_{mp}(\rho,\rho)\nn\\
&=&-\frac\partial{\partial\alpha}\frac1{4\pi}\int_0^\infty d\kappa\,\kappa 
\,J_0(\kappa\delta)\sum_{m=1}^\infty
\ln\Delta_\nu(\kappa a,\kappa b).\label{regtorque}
\eea
Is this the negative derivative of the energy?

The expectation value of the energy density, given in Eq.~(\ref{veve}),
leads to the following expression for the energy per unit length, using the same point
splitting as above, instead of Eq.~(\ref{unregenergy}):
\be
\mathcal{E}=-\int_0^\infty \frac{d\kappa\,\kappa^3}{2\pi}f(\kappa\delta,\phi)\sum_{m=1}^\infty
\int_a^b d\rho\,\rho\,g_\nu(\rho,\rho),
\ee
where the regulator function is
\bea
f(\kappa\delta,\phi)&=&\int_0^{2\pi}\frac{d\gamma}{2\pi}\cos^2\gamma \,
e^{i\kappa\delta\cos(\gamma-\phi)}\nn\\
&=&\left[\frac12 J_0(\kappa\delta)-\frac1{\kappa\delta}J_1(\kappa\delta)\right]
\cos2\phi
+\frac12 J_0(\kappa\delta),\nn\\
\label{regulator}
\eea
which equals 1/2 as $\delta\to 0$.  
As before, the radial integral is
\be
\int_a^b d\rho\,\rho\, g_\nu(\rho,\rho)=\frac1{2\kappa}
\frac\partial{\partial\kappa}\ln\Delta_\nu(\kappa a,\kappa b),
\ee
so if we integrate by parts,\footnote{
Integration by parts is formally legitimate for $\phi=\pi/2$, that is,
$z$-splitting, in that then the surface term at infinity, although 
divergent, is a constant in $\alpha$.  But even that is not so for 
$\phi=0$, $t$-splitting.  Whether this fact has any relation to the 
$t$-splitting anomaly discussed in the next section is not clear.}
the energy/length is
\be
\mathcal{E}=\frac1{4\pi}\int_0^\infty d\kappa\frac\partial{\partial\kappa}
\left[\kappa^2f(\kappa\delta,\phi)
\right]\sum_{m=1}^\infty\ln\Delta_\nu(\kappa a,\kappa b).\label{eibp}
\ee
Is this the quantity appearing in Eq.~(\ref{regtorque})?  This would seem to
require that
\be
\frac\partial{\partial\kappa}\left[\kappa^2 f(\kappa\delta,\phi)\right]
=\kappa J_0(\kappa\delta).\label{neutralcond}
\ee
Indeed this is so for the $z$ regulator, $\phi=\pi/2$. 
But it is not so for time splitting, where
\be
\frac\partial{\partial\kappa}\left[\kappa^2 f(\kappa\delta,0)\right]=
\kappa J_0(\kappa\delta)+\kappa^2\delta J_0'(\kappa\delta).
\ee

We will now proceed to see how the divergent and finite parts of the energy
behave as functions of $\delta$, $\phi$, and $\alpha$.

\section{Weyl Terms}
\subsection{Leading Divergences}
Let us now extract the leading divergences, as the cutoff $\delta\to0$, in the
energy using the form (\ref{eibp}) (before the suspect integration by parts),
\be
\mathcal{E}=-\frac1{4\pi}\int_0^\infty d\kappa \,\kappa^2 f(\kappa\delta,\phi)
\sum_{m=1}^\infty
\frac\partial{\partial\kappa}\ln\Delta_\nu(\kappa a,\kappa b).
\label{regulatede}
\ee
To do so, we use the uniform asymptotic expansions 
\cite[10.41.ii]{nist} 
of the modified Bessel functions,
applicable as $\nu\to\infty$:
\begin{subequations}
\label{asymik}
 \bea
I_\nu(\nu z)&\sim&\sqrt{\frac{t}{2\pi\nu}}e^{\nu\eta}\left(1+
\sum_{k=1}^\infty\frac{u_k(t)}{\nu^k}\right),\\
K_\nu(\nu z)&\sim&\sqrt{\frac{\pi t}{2\nu}}
e^{-\nu\eta}\left(1+\sum_{k=1}^\infty\frac{(-1)^ku_k(t)}{\nu^k}\right),
\eea
\end{subequations}
where $t=(1+z^2)^{-1/2}$, $\eta(z)=1/t+\ln\frac{z}{1+1/t}$.  It follows that
\be
\frac{d\eta}{dz}=\frac1{zt}.
\ee
Here the $u_k(t)$ are polynomials in $t$ of degree $3k$.  If we 
retain only the leading factor, we get, with $\nu z=\kappa b$,
\be
\Delta_\nu(\nu z a/b,\nu z)\sim \frac{
\sqrt{t(z)t(z a/b)}}{\nu}\sinh \nu[\eta(z)-\eta(za/b)],\label{lf}
\ee
where $\nu z = \kappa b$.
Thus, the leading divergence is obtained from
\be
\ln\Delta_\nu(\kappa a,\kappa b)\sim \nu[\eta(z)-\eta(z a/b)],
\ee
which has derivative
\be
\frac\partial{\partial\kappa}\ln\Delta_\nu\sim\frac{b}z
\left[\sqrt{1+z^2}-\sqrt{1+(za/b)^2}\right].
\ee

For the $z$-point-splitting, $\phi=\pi/2$, we have the following 
expression for the most divergent contribution:
\bea
\mathcal{E}_4(\pi/2)&=& -\frac{1}{4\pi b\delta}\sum_{m=1}^\infty 
\nu^2\int_0^\infty dz\,J_1(\nu z \delta/b)\nn\\
&&\quad\times\left[\sqrt{1+z^2}-\sqrt{1+(za/b)^2}\right].\label{e4pi2}
\eea
Although not classically convergent, the $z$ integral has a well-defined 
meaning:
\be
\int_0^\infty dz\,J_1(az)\sqrt{1+z^2}=\frac1a+\frac1{a^2}e^{-a},
\ee
and then, approximating the sum on $m$ by an integral:
\be
\sum_{m=1}^\infty\to\frac\alpha\pi\int_0^\infty d\nu,\label{sumtoint}
\ee
we immediately find
\bea
\mathcal{E}_4(\pi/2)&\sim&-\frac1{4\pi \delta^3}\frac\alpha\pi\int_0^\infty 
d\nu\left[be^{-\nu \delta/b}
-a e^{-\nu \delta/a}\right]\nn\\
&=&-\frac{\alpha}{4\pi^2\delta^4}(b^2-a^2)=-\frac{A}{2\pi^2 \delta^4}
=\mathcal{E}^{(4)}(\pi/2),\nn\\
\label{zco}
\eea
which exhibits the expected quartic divergence as the cutoff $\delta$ tends to zero.
Here the area of the annular region is $A=\alpha(b^2-a^2)/2$.

If we use the time-splitting cutoff instead, so $\phi=0$, and the regulator function is
\be
f(\kappa\delta,0)=J_0(\kappa\delta)-\frac1{\kappa\delta}J_1(\kappa\delta),
\label{f0}
\ee
we see that the second term gives the negative of the result (\ref{zco}),
while the first  involves the integral
\be
\int_0^\infty dz\,z\sqrt{1+z^2}J_0(za)=-\frac{1+a}{a^3}e^{-a},
\ee
which involves an integration by parts, where we omit the term at infinity.
Following the same procedure as in the preceding paragraph we are led to the sum
of the contributions from the first and second terms in the cutoff function 
(\ref{f0}):
\be
\mathcal{E}_4(0)=\frac{A}{\pi^2\delta^4}
+\frac{A}{2\pi^2\delta^4}=\frac{3A}{2\pi^2\delta^4}
=\mathcal{E}^{(4)}(0),\label{tco}
\ee
which is exactly the expected Weyl divergence with a temporal cutoff 
\cite{Fulling:2003zx,Abalo:2010ah,Abalo:2012jz}.
It might seem remarkable
that not only are the coefficients different in the two regularization schemes,
but even the signs are reversed.  However, as we will see at the end of this
section, this is entirely to be expected.

Because of the linear dependence of the area of the annular region
on the wedge angle $\alpha$, it is apparent that the leading Weyl
term contributes to the torque.  Only the cutoff in the ``neutral''
$z$ direction is consistent with the torque according to 
Eqs.~(\ref{regtorque}), (\ref{eibp}), and (\ref{neutralcond}).

The temporal cutoff indeed exhibits an anomaly.

A word about notation.  Subscripts on energy terms refer to the order of the
term in the uniform asymptotic expansion, so, for example, $\mathcal{E}_4$
refers to the leading, exponential behavior, while superscripts within
parentheses refer to the order in powers of $\delta^{-1}$.  Thus, as we
shall see,  
$\mathcal{E}_4$ contains not only all of $\mathcal{E}^{(4)}$, but part of
$\mathcal{E}^{(3)}$.

\subsection{Subleading divergences}
Now we wish to extract all the divergent terms in the energy.  First, we
note that the leading term in the energy, $\mathcal{E}_4$, was not calculated
exactly, because of the approximation (\ref{sumtoint}).  Instead, we can carry
out the sum exactly, 
\be
\sum_{m=1}^\infty e^{-m t}=\frac1{e^t-1}=\sum_{n=0}^\infty B_n \frac{t^{n-1}}{n!},\label{bernoulli}
\ee
where the expansion is valid for small $t$, and $B_n$ is the $n$th Bernoulli
number.
We can evaluate the sum on $m$ appearing in 
Eq.~(\ref{e4pi2}) exactly, and thus determine
the behavior of $\mathcal{E}_4$ for small $\delta$:
\begin{subequations}
 \bea
\mathcal{E}_4(\pi/2)&=&-\frac{\alpha(b^2-a^2)}{4\pi^2\delta^4}+
\frac{b-a}{8\pi\delta^3},\label{e4z}\\
\mathcal{E}_4(0)&=&\frac{3\alpha(b^2-a^2)}{4\pi^2\delta^4}-
\frac{b-a}{4\pi\delta^3}.\label{e4t}
\eea
\end{subequations}
The $\delta^{-4}$ terms coincide with those in 
Eqs.~(\ref{zco}) and (\ref{tco}).
The corrections to this evaluation are finite as $\delta\to 0$.

The next subleading term comes from the square-root factor in Eq.~(\ref{lf}),
where the contributing part of $\ln\Delta_\nu$  is
\be
\frac12 \ln t(z)+\frac12\ln t(za/b)=-\frac14\ln(1+z^2)-
\frac14\ln(1+z^2a^2/b^2),
\ee
so for the $\phi=\pi/2$ cutoff
\bea
\mathcal{E}_3(\pi/2)&=&\frac1{8\pi \delta }\sum_{m=1}^\infty \nu
\bigg\{\frac1b\int_0^\infty dz\frac{z^2}{1+z^2}J_1(\nu z\delta/b)\nn\\
&&\qquad\mbox{}+(b\to a)
\bigg\}.
\eea
The integrals here converge in the Fresnel sense:
\be
\int_0^\infty dz\frac{z^2}{1+z^2}J_1(\nu z\delta/b)=K_1(\nu\delta/b),
\ee
so this term in the energy is
\be
\mathcal{E}_3=\frac1{8\pi\delta}\sum_{m=1}^\infty \nu\left[\frac1b K_1(\nu\delta/b)
+\frac1a K_1(\nu\delta/a)\right].\label{k1sum}
\ee
Again, in the first approximation, we may replace the sum by an integral,
so we find approximately
\be
\mathcal{E}_3(\pi/2)\sim\frac{\alpha(a+b)}{16\pi\delta^3}.
\ee
However, there is a subleading divergent contribution, 
as we can see by carrying out the $m$ sum exactly, using \cite{GR}
\bea
&&\sum_{m=1}^\infty K_0(mx)\cos mxt=
\frac12\left(\gamma+\ln\frac{x}{4\pi}\right)
+\frac{\pi}{2x\sqrt{1+t^2}}\nn\\
&&\mbox{}+\frac\pi2\sum_{l=1}^\infty\left\{\frac1{\sqrt{x^2+(2l\pi-tx)^2}}-\frac1{2l\pi}\right\}\nn\\
&&\mbox{}+\frac\pi2\sum_{l=1}^\infty\left\{\frac1{\sqrt{x^2+(2l\pi+tx)^2}}-\frac1{2l\pi}\right\}.\label{grsum}
\eea
Since $K_0'(x)=-K_1(x)$, this implies (this result can also be
derived from Eq.~(2.7) of Ref.~\cite{Kirsten:1992fv})
\be
\sum_{m=1}^\infty m K_1(m\epsilon)\sim\frac\pi{2\epsilon^2}-\frac1{2\epsilon},
\ee
as $\epsilon\to0$ with no further corrections.  When this is inserted into
Eq.~(\ref{k1sum}) we obtain
\be
\mathcal{E}_3(\pi/2)=\frac{\alpha(a+b)}{16\pi\delta^3}
-\frac1{8\pi\delta^2}.\label{e3p2}
\ee
Now note that when the $O(\delta^{-3})$ contribution is combined with the $O(\delta^{-3})$
term in Eq.~(\ref{e4z}) we obtain
\be
\mathcal{E}^{(3)}(\pi/2)=\frac{P}{16\pi\delta^3},
\ee
in terms of the perimeter of the annular region,
\be
P=\alpha(a+b)+2(b-a).
\ee
The $O(\delta^{-2})$ correction is the expected corner divergence,
\be
\mathcal{E}^{(2)}(\pi/2)=-\frac{C}{48\pi\delta^2},\quad C=
4\left(\frac\pi{\pi/2}-\frac{\pi/2}\pi\right)=6.
\label{cornerdiv}
\ee

For time-splitting regularization, $\phi=0$, we subtract this result
from that with $J_1(\kappa\delta)\to \kappa\delta J_0(\kappa\delta)$
[recall Eq.~(\ref{f0})], where the latter involves the integral
\be
\int_0^\infty dz \frac{z^3}{z^2+1}J_0(z\nu\delta/b)=-K_0(\nu\delta/b).
\ee
Thus this term is 
\be
\mathcal{E}'_3=-\frac1{8\pi}\sum_{m=1}^\infty \nu^2\left[\frac1{b^2}K_0(\nu\delta/b)
+\frac1{a^2}K_0(\nu\delta/a)\right].
\ee
Once again, the sum can be carried out exactly using Eq.~(\ref{grsum}),
\be
\sum_{m=1}^\infty m^2 K_0(m\epsilon)\sim\frac\pi{2\epsilon^3}
\ee
as $\epsilon\to 0$.  Thus only the leading term is divergent here:
\be
\mathcal{E}'_3=-\frac{\alpha(b+a)}{8\pi\delta^3},
\ee
and then subtracting the $\phi=\pi/2$ result (\ref{e3p2})  gives the divergent term
for time splitting,
\be
\mathcal{E}_3(0)=-\frac{\alpha(a+b)}{16\pi\delta^3}+\frac1{8\pi\delta^2}.
\ee
Thus we obtain the expected perimeter divergence \cite{Abalo:2010ah}
\be
\mathcal{E}^{(3)}(0)=-\frac{P}{8\pi\delta^3},
\ee
and the expected corner divergence:
\be
\mathcal{E}^{(2)}(0)=\frac{C}{48\pi\delta^2},\quad C=6.
\ee

The next correction comes from the expansion of the logarithm of $\Delta_\nu$
of the terms involving an expansion in powers of $1/\nu$ seen in Eq.~(\ref{asymik}).  That gives
\bea
\frac{d}{d\kappa}\frac1\nu\left[u_1(t)-u_1(\tilde t)\right]&=&
-\frac{b}{\nu^2}\frac{z}8\bigg[t^3-5t^5\nn\\
&&\quad\mbox{}-\frac{a^2}{b^2}(\tilde t^3-5\tilde t^5)
\bigg],
\eea
with $\tilde t=t(za/b)$.
This gives for the energy contribution for the $z$ splitting
\bea
\mathcal{E}_2(\pi/2)&=&\frac1{32\pi\delta}\sum_{m=1}^\infty \bigg\{\frac{1}b\int_0^\infty
dz\,z^2 \,J_1(\nu z\delta/b)t^3(1-5t^2)\nn\\
&&\qquad\mbox{}-(b\to a)\bigg\}.
\eea
The integrals on $z$ are readily evaluated as
\begin{subequations}
 \bea
\int_0^\infty dz\,z^2 t^3\,J_1(z c)&=&e^{-c},\\
\int_0^\infty dz\,z^2 t^5\,J_1(z c)&=&\frac{c}3e^{-c},
\eea
\end{subequations}
and then the sum on $m$ can be evaluated, and approximated by
the leading terms in the Bernoulli expansion (\ref{bernoulli}), resulting in a term,
\be
\mathcal{E}_2(\pi/2)=\frac1{64\pi\delta}\left(\frac1a-\frac1b\right).\label{e2}
\ee
Here a cancellation of the term of order $\delta^{-2}$ has occurred.  
It is expected that the curvature term of this order should cancel, 
because it should be proportional to
\be
\oint_{\partial V} dS\,K=0,
\ee
because the curvature $K$ is $-1/b$ or $1/a$ for the outer and inner arc, 
respectively, so the two contributions cancel.  Recall that we have
already encountered, at the order $\delta^{-2}$,  a corner
term proportional to $C=6$ as seen in 
Eq.~(\ref{cornerdiv}).
Before discussing the meaning of the remaining term in Eq.~(\ref{e2}),
we give the result for the time-splitting regularization, that is, $\phi=0$, 
which is found in just the same way:
\be
\mathcal{E}_2(0)=\frac1{64\pi\delta}\left(\frac1a-\frac1b\right)
-\frac1{64\pi\delta}\left(\frac1a-\frac1b\right)=0.\label{e2t}
\ee
The cancellation indicated occurs between the $J_0$ and $J_1$ regulator terms
occurring in Eq.~(\ref{f0}).
Thus this term, which is a corner curvature correction \cite{Dowker:1995sp},
is not present for the temporal cutoff.

Penultimately, we extract the  divergent contribution coming from the $1/\nu^2$
contributions to the logarithm:
\bea
&&\ln\left(1+\frac{u_1}\nu+\frac{u_2}{\nu^2}+\dots\right)
\left(1-\frac{\tilde u_1}\nu+\frac{\tilde u_2}{\nu^2}+\dots\right)\nn\\
&=&\frac1\nu\frac{3t-5t^3-(t\to\tilde t)}{24}\nn\\
&&\quad\mbox{}+\frac1{\nu^2}\frac{t^2(1-t^2)(1-5t^2)+
(t\to\tilde t)}{16}+\dots,\label{logasym}
\eea
where 
$\tilde u_n=u_n(\tilde t)$. Then the form of
this energy contribution is, for the $z$ point-splitting,
\bea
\mathcal{E}_1(\pi/2)&=&\frac1{32\pi\delta}\sum_{m=1}^\infty \frac1\nu\bigg\{\frac1b
\int_0^\infty dz\,z^2 J_1(\nu z\delta/b) t^4\nn\\
&&\quad\times(1-12t^2+15t^4)+(b\to a)\bigg\}.
\eea
Now it is easiest to sum on $m$ first, yielding approximately,
\be
\sum_{m=1}^\infty \frac1\nu J_1(\nu z\delta/b)\sim \frac\alpha\pi
\int_0^\infty \frac{d\nu}\nu J_1(\nu z\delta/b)=\frac\alpha\pi,
\ee
but actually, the sum can be done exactly,
\be
\sum_{m=1}^\infty \frac1m J_1(m x)=1-\frac{x}4.
\ee
The second term here gives rise to a logarithmically divergent
$z$ integral, which we represent by $\ln(1/\delta)$.
The remaining $z$ integral is
elementary, and we find
\be
\mathcal{E}_1(\pi/2)=-\frac\alpha{1024\pi\delta}\left(\frac1b+\frac1a\right)
+\frac1{128\pi}\left(\frac1{a^2}+\frac1{b^2}\right)\ln\delta.\label{e1p2}
\ee
This evaluation can also be carried out by doing the $z$ integral first,
and then the sum on $m$, which involves the additional asymptotic
summation formula
\be
\sum_{m=1}^\infty m^2 K_2 (m\epsilon)\sim \frac{3\pi}{2\epsilon^3}
-\frac1{\epsilon^2}.
\ee
The geometric quantity appearing in the first term of Eq.~(\ref{e1p2})
is that expected for the surface integral of
the square of the curvature over the two arcs.  The second term is again a 
curvature corner correction. When the same
calculation is carried out for time-splitting, we again find that the $J_0$
term cancels the negative of the $\delta^{-1}$  term, while the $\ln\delta$
term does not change:
\be
\phi=0:\quad \mathcal{E}_1(0)=
\frac1{128\pi}\left(\frac1{a^2}+\frac1{b^2}\right)\ln\delta.
\ee

Finally, we extract the $O(\nu^{-3})$ contribution from the logarithm 
 (\ref{logasym}),
\be
\ln(\cdots)\sim\dots \frac{375 t^3-4779t^5+9945 t^7-5525 t^9}{5760\nu^3}
 -(t\to \tilde t).
\ee
When this is inserted into the formula for the energy, we encounter the 
integrals
\be
\int_0^\infty dz\,z^2 J_1(z a)t^{2n+1}=
\left(\frac{a}2\right)^{n-1/2}\frac{K_{n-3/2}(a)}{\Gamma(n+1/2)}.
\ee 
To get the divergent term, the sum over $m$  may be  approximated 
by an integral, with the result
\be
\mathcal{E}_0\sim\frac{\ln\delta}{1260\pi^2}\alpha\left(\frac1{b^2}
-\frac1{a^2}\right).
\ee
The same result applies for either $\phi=0$ or $\pi/2$ regularization 
(see below).
This is the expected integrated curvature-cubed divergence.

\subsection{Conclusions} 

Let us summarize by giving the divergent contribution for the 
$z$ and $t$ regularizations:
\begin{subequations}
\label{divergences}
 \bea
\mathcal{E}_{\rm div}(\pi/2)&=&-\frac{A}{2\pi^2 \delta^4}+
\frac{P}{16\pi\delta^3}-\frac{C}{48\pi\delta^2}\nn\\
&&\mbox{}+\frac1{64\pi\delta}\left(\frac1a-\frac1b\right)
-\frac\alpha{1024\pi\delta}\left(\frac1{a}+\frac1b\right)\nn\\
&&\mbox{}+\frac{\ln\delta/\mu}{128\pi}\left(\frac1{a^2}+\frac1{b^2}\right)\nn\\
&&\mbox{}+\frac{\ln\delta/\mu}{1260\pi^2}\alpha\left(\frac1{b^2}-\frac1{a^2}
\right),\label{divergence0}\\
\mathcal{E}_{\rm div}(0)&=&\frac{3A}{2\pi^2 \delta^4}-\frac{P}{8\pi\delta^3}+
\frac{C}{48\pi\delta^2}\nn\\
&&\mbox{}+\frac{\ln\delta/\mu}{128\pi}\left(\frac1{a^2}+\frac1{b^2}\right)\nn\\
&&\mbox{}+\frac{\ln\delta/\mu}{1260\pi^2}\alpha\left(\frac1{b^2}-\frac1{a^2}
\right).
\eea
\end{subequations}
Here we have inserted an arbitrary scale $\mu$ into the logarithm,
which will lead to an arbitrariness in the finite part, as we shall see
in the next section.
The absence of the $\delta^{-1}$ term in the time-splitting
scheme is exactly that observed, for example, in Ref.~\cite{Fulling:2003zx}.
Moreover, the ratio of coefficients for the $\delta^{-n}$ terms, $n=4, 3, 2,
1, 0$, in the two schemes is exactly that found in Ref.~\cite{Estrada:2012yn},
namely, $-3$, $-2$, $-1$, $0$ and $1$, that is, for both divergent (as 
$\delta\to0$)
and finite contributions,
\be
\mathcal{E}(0)=\frac{d}{d\delta}\left[\delta 
\mathcal{E}(\pi/2)\right],\label{rele0epi2}
\ee which follows immediately from
Eq.~(\ref{regulatede}) and the recursion relation
\be
J_1'(z)=J_0(z)-\frac1z J_1(z).
\ee

The coefficients found in this section are in agreement with the 
calculations of 
Dowker and Apps \cite{Dowker:1995sp,apps}
and  Nesterenko, Pirozhenko,
and Dittrich \cite{Nesterenko:2002ng} of the heat kernel 
coefficients for
a wedge intercut with a single coaxial circular cylinder, from which the
above divergences, with exactly the coefficients found, can be  
inferred by the formulas of~Ref.~\cite{Fulling:2003zx}, relating the heat
kernel \cite{kirsten,gilkey} to the cylinder kernel \cite{lukosz,Bender:1976wb}.
The trace of the cylinder kernel $T(t)$ is defined in terms of the 
eigenvalues of the Laplacian in $d$ dimensions,
\be
T(t)=\sum_j e^{-\lambda_j t}\sim \sum_{s=0}^\infty e_s t^{s-d}+\sum_{s=d+1\atop
s-d {\rm odd}}f_s t^{s-d}\ln t,
\ee
where the expansion holds as $t\to0$ through positive values.  The
energy is given by
\be
E(t)=-\frac12\frac\partial{\partial t}T(t),
\ee
which corresponds to the energy computed here with $\phi=0$, that is,
time-splitting.  In view of the relation (\ref{rele0epi2}) 
between $z$ and $t$ splitting, we see that the $z$-splitting
energy should be identical to the expansion
 of $-\frac1{2t} T(t)$ with $t\to\delta$.
In this way we transcribe the results of Ref.~\cite{Nesterenko:2002ng} for
the traced cylinder kernel per unit length:
\bea
-\frac1{2t}T(t)&\sim&-\frac{A}{2\pi^2 t^4}+\frac{P}{16\pi t^3}
-\frac{1}{16\pi^2 t^2}+\frac{1-\alpha/16}{64\pi R t}\nn\\
&&\quad\mbox{}+\frac{\ln t}{4\pi^2R^2}
\left(\frac{\pi}{32}-\frac{\alpha}{315}\right).
\eea
This exactly agrees with our result when $a\to R$ and $b\to \infty$
(except in the first two terms).  The reason for the factor of 2 discrepancy
in the third (corner) term is that Nesterenko et al.\ have only two corners,
not four.

\section{Finite Part of Energy}
To extract the finite part of the interior energy of the annular region,
we have to compute first the finite parts resulting from the asymptotic terms
that gave rise to the divergences (\ref{divergences}).  These are easily
worked out:
\begin{subequations}\label{finiteparts}
\bea
\mathcal{E}^f_4&=&-\frac{\pi^2}{2880}\frac1{\alpha^3}\left(\frac1{a^2}-
\frac1{b^2}\right),\\
\mathcal{E}_3^f&=&\frac{\zeta(3)}{64\pi}\frac1{\alpha^2}\left(\frac1{a^2}+
\frac1{b^2}\right),\\
\mathcal{E}_2^f&=&-\frac1{144}\frac1\alpha\left(\frac1{a^2}-\frac1{b^2}\right),
\\
\mathcal{E}^f_1&=&\frac1{128\pi}\left(\gamma+\frac74-\ln4
+\ln\frac{\mu}{b\alpha}
\right)\frac1{b^2}\nn\\
&&\quad\mbox{}+(b\to a),\\
\mathcal{E}^f_0&=&\frac\alpha{1260\pi^2}\left(
\ln\frac{\mu}{b\alpha}+\ln\pi
-\frac{397}{48}\right)\frac1{b^2}\nn\\
&&\qquad\mbox{}-(b\to a).
\eea
\end{subequations}
These results are valid for any regularization scheme, except for the last
two terms, where the application of the operator in Eq.~(\ref{rele0epi2}) 
gives rise to additional terms from the logarithmically divergent terms in 
Eq.~(\ref{divergence0}):
\bea
\mathcal{E}^f(0)&=&\mathcal{E}^f(\pi/2)+\frac1{128\pi}\left(\frac1{a^2}
+\frac1{b^2}\right)\nn\\
&&\quad\mbox{}-\frac\alpha{1260\pi^2}\left(\frac1{b^2}-\frac1{a^2}
\right).
\eea
In fact, the appearance of $\mu$ in the logarithm means that the energy
is ambiguous up to a linear term in $\alpha$:
\be
\mathcal{E}\to\mathcal{E}+A+B\alpha.\label{linear}
\ee
We will determine the constants $A$ and $B$ by requiring that the energy
approach zero for large enough angular separation between the radial planes in
the annulus.
 This precisely means that constant and linear terms in $\alpha$ in 
the energy are eliminated.  Because all the divergent terms seen in 
Eq.~(\ref{divergences}) are of this form, this means that this 
renormalization process
will remove all divergences, and then since the resulting energy is finite,
Eq.~(\ref{tauenergy}), as expected, will hold.

The finite energy is the sum of these terms in Eq.~(\ref{finiteparts}), 
plus the remainder,
which comes from subtracting these asymptotic contributions to $\ln\Delta_\nu$
from the original expression in Eq.~(\ref{regulatede}).  Because this
remainder is finite, we can replace the regulator function by 1/2, and so
\be
\mathcal{E}_R=-\frac1{8\pi}\int_0^\infty d\kappa\,\kappa^2\sum_{m=1}^\infty
\frac{\partial}{\partial \kappa}\left[\ln\Delta_\nu-\sum_{n=4}^0
\ln\Delta_\nu^{(n)}\right].
\ee 
Let us write these in terms of the natural variables for the uniform asymptotic
expansion, $\nu=m \pi/\alpha$ and  $z=\kappa b/\nu$:
\bea
\mathcal{E}_R&=&-\frac1{8\pi b^2}\sum_{m=1}^\infty \nu^3\int_0^\infty d z\,z^2
\bigg\{f(\nu,z,a/b)\nn\\
&&\qquad\qquad\mbox{}+\sum_{n=4}^0 f_n(\nu,z,a/b)\bigg\},\label{remainder}
\eea
where with $I=I_\nu(\nu z)$, $\tilde I=I_\nu(\nu z a/b)$, $K=K_\nu(\nu z)$,
$\tilde K=K_\nu(\nu z a/b)$,
the original integrand is 
\be
f(\nu, z, a/b)=\frac{(I'\tilde K-\tilde I K')+\frac{a}b (I\tilde K'-\tilde I'
K)}{I\tilde K-\tilde I K}.
\ee
The subtractions of the asymptotic terms give
\begin{subequations}
\bea
f_4&=&-\frac1{zt}+\frac{a}b\frac1{\tilde z\tilde t},\\
f_3&=&\frac1{2\nu}\left(zt^2+\frac{a}b \tilde z\tilde t^2\right),\\
f_2&=&\frac1{8\nu^2}\left[z(t^3-5 t^5)-\frac{a}b\tilde z(\tilde t^3
-5\tilde t^5)\right],\\
f_1&=&\frac1{8\nu^3}\left[z(t^4-12t^6+15t^8)+\frac{a}b\tilde z(\tilde t^4
-12\tilde t^6+15 \tilde t^8)\right],\nn\\
\\
f_0&=&\frac1{5760\nu^4}\bigg[z(1125 t^5-23895 t^7+69615 t^9-49725 t^{11})\nn\\
&&\quad\mbox{}-\frac{a}b\tilde z(1125 \tilde t^5-23895 \tilde t^7
+69615 \tilde t^9-49725 \tilde t^{11})\bigg].\nn\\
\eea
\end{subequations}

To improve convergence, we should subtract off two more 
(finite) aysmptotic terms,
and add back in the corresponding finite terms:
\begin{subequations}
\bea
\mathcal{E}_{-1}&=&-\frac{\alpha^2}{6144\pi}\left(\frac1{a^2}+\frac1{b^2}
\right),\\
\mathcal{E}_{-2}&=&-\frac{29}{180180}\frac{\zeta(3)}{\pi^4}\alpha^3
\left(\frac1{a^2}-\frac1{b^2}\right).
\eea
\end{subequations}
The corresponding subtractions that should be added to the integrand in
Eq.~(\ref{remainder}) are
\begin{subequations}
\bea
f_{-1}&=&\frac1{128\nu^5}\bigg[z(52 t^6-1704 t^8+8496 t^{10}-13560 t^{12}\nn\\
&&\quad\mbox{}+6780 t^{14})+\frac{a}b(z\to\tilde z)\bigg],\\
f_{-2}&=&\frac1{322560\nu^6}\bigg[z(337995 t^7-15765435 t^9\nn\\
&&\quad\mbox{}+117697230 t^{11}-311150070 t^{13}+339168375 t^{15}\nn\\
&&\quad\mbox{}-130449375 t^{17})- \frac{a}b(z\to\tilde z)\bigg].
\eea
\end{subequations}

\begin{figure}
\begin{center}
\includegraphics[scale=.7]{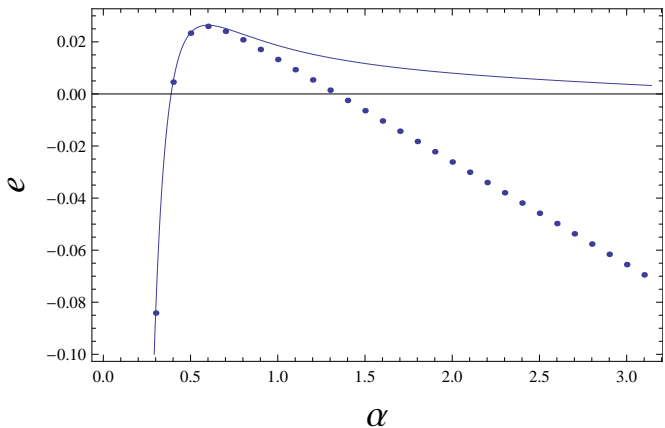}
\end{center}
\caption{\label{fig:wander} The energy of a finite annular sector,
with ratio of inner and outer annular radii $a/b=0.5$, as a function
of the wedge angle $\alpha$.  The solid curve shows the explicit angular
dependence resulting from the subtractions, $w(\alpha)$, while the data
points are the total energy obtained 
by adding $w$ to the numerical integration 
of the remainder $e_R(\alpha)$.  It will be noted that below the maximum,
the total energy is indistinguishable from that obtained from $w$, while
above the maximum, the total energy is asymptotically linear, and deviates
significantly from $w$.}
\end{figure}

\begin{figure}
\begin{center}
\includegraphics[scale=.7]{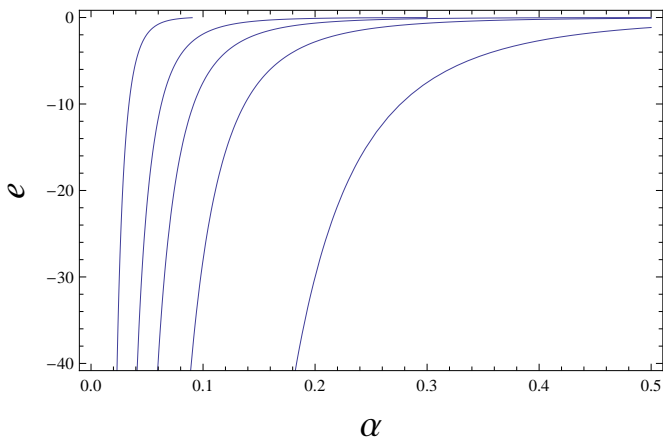}
\end{center}
\caption{\label{fig:renormen} The renormalized energy for the annular sector,
obtained by taking the energy $\mathcal{E}^f$ and subtracting the linear
dependence occuring for sufficiently large angles.  Shown, from left to right,
are energies for ratios of inner and outer radii, $a/b$, equal to 0.9, 0.7,
0.5, 0.3, and 0.1, respectively.}
\end{figure}

Let us write
\be
\mathcal{E}^f=\frac1{b^2}\left[w(\alpha,a/b)+e_R(\alpha,a/b)\right],
\ee
where $e_R=\mathcal{E}_R b^2$ and
\be
w=b^2\sum_{n=4}^{-2}\mathcal{E}_n^f.
\ee
  A typical example of the behavior of
$w$, coming from the explicit subtractions, and the remainder $e_R$ is
shown in Fig.~\ref{fig:wander}.
The qualitative  features hold in every case.  
The energy is accurately given by $w$
below the point at which the energy reaches a maximum, and above that point,
the total energy, which now deviates signficantly from $w$, is very accurately
linear. 
Therefore, to obtain the finite, renormalized energy, we subtract
from the energy a fit to this dependence, in accordance with the remarks
following Eq.~(\ref{linear}).  In this way we obtain the results
shown in Fig.~\ref{fig:renormen}.
The energy, the negative gradient of which now unambiguously is the torque
on one of the radial planes, is large and negative for small wedge angle 
$\alpha$ and then rapidly tends to zero.  As the ratio $a/b$  gets smaller
the region where the attractive torque is significant gets larger.  These
curves are very similar to those found for an annular piston in 
Ref.~\cite{Milton:2009bz}, the difference there being that both sides of
the radial walls are considered, so $\mathcal{E}(\alpha)\to 
\mathcal{E}(\alpha)+\mathcal{E}(2\pi-\alpha)$, so a similar attraction appears
near $\alpha=2\pi$.  Indeed, when a smaller plot range is specified the results
agree rather closely with those found in Ref.~\cite{Milton:2009bz}.
There, only three-body (wedge-inner cylinder-outer cylinder) effects were considered 
from the outset.  By our process of renormalization we have removed two-body and one-body terms, 
such as the energy due to the wedge by itself.

\section{Discussion}
The divergent terms, the pressure anomaly related to the direction of
point-splitting, and
the ambiguity associated with logarithmic divergences,  are present in the
energy as linear terms in the wedge angle $\alpha$.  Constant terms in
$\alpha$, of course, do not contribute to the torque, and linear terms yield
a constant torque, that is, one
 independent of the wedge angle.  Any such constant torque 
has no physical significance and should be 
subtracted. Indeed, because of the logarithmic divergence associated
with curvature, any linear dependence in the wedge angle is ambiguous,
and that dependence must be determined by a physical requirement.  Here 
that requirement is supplied by the condition that the energy must vanish
for sufficiently large wedge angles. 
Furthermore, any linear dependence in $\alpha$ would be cancelled if the
exterior region of the annular piston were included, for which $\alpha\to
2\pi-\alpha$.  For this reason,  only the finite, unambiguous,
nonlinear dependence in the energy has physical significance.
The torque anomaly that appeared for $t$-splitting occurs only in the
divergent terms and therefore is removed by the process of renormalization.

Finally, let us make a remark about the situation when there is no inner
boundary, so the sector consists of the region $0\le\rho<b$, $0<\theta<\alpha$.
Why could one not proceed as here for the inner radius $a>0$?  Then in the
divergent terms we would have for the corner divergence
\be
\mathcal{E}_{\rm corner\, div}=\mp \frac{C}{48\pi\delta^2},
\ee
for $z$- and $t$-splitting respectively.  Now the corner coefficient
contains the apex:
\be
C_{\rm apex}=\frac\pi\alpha-\frac\alpha\pi.\ee
Now, the divergent terms have a nonlinear dependence on $\alpha$, rendering
it impossible to extract a finite energy through
``renormalization.''  This irreducible singularity
presumably is the mirror of the torque anomaly of Ref.~\cite{fulling12}.

In subsequent papers we will explore the electromagnetic situation, with
perfectly conducting boundary conditions, and how these results can be 
understood from a local analysis of the stress tensor.

\acknowledgments
We thank the U.S. National Science Foundation and the Julian Schwinger
Foundation for the support of this research.  We thank our many collaborators,
especially Iver Brevik, Stuart Dowker, Stephen Holleman,
K. V. Shajesh, and Jef Wagner,  for helpful discussions.

\end{document}